\documentclass{appolb}
\usepackage{amsmath,amssymb,amsfonts,amsthm}

\newcommand{\com}[2]{[#1,#2]}
\newcommand{\starcom}[2]{[#1\stackrel{\star}{,}#2]}
\newcommand{\pair}[2]{\langle #1,#2\rangle}
\def\Nabla{\triangledown}

\title{%
  Symmetry~Reduction and Exact Solutions in Twisted~Noncommutative~Gravity}

\author{%
  Alexander Schenkel%
    \thanks{e-mail: \texttt{aschenkel@physik.uni-wuerzburg.de}}\\
  \hfil\\
  Institut f\"ur Theoretische Physik und Astrophysik\\
  Universit\"at W\"urzburg, Am Hubland, 97074 W\"urzburg, Germany}

\date{\today}

\begin{document}
\pagestyle{plain}
\maketitle

\begin{abstract}
We review the noncommutative gravity of Wess et al.~\cite{Aschieri:2005yw,Aschieri:2005zs}
and discuss its physical applications.
We define noncommutative symmetry reduction and construct deformed symmetric solutions of
 the noncommutative Einstein equations. We apply our framework to find explicit deformed cosmological and black hole 
solutions and discuss their phenomenology.

This article is based on a joint work with Thorsten Ohl \cite{Ohl:2008tw,Ohl:2009pv}.
\end{abstract}


\section{Introduction}
Despite the great success of Einstein's general theory of relativity,
it is generally believed that it has to be modified at small
distances, incorporating quantum effects of spacetime. To achieve this
goal and arrive at a consistent theory of quantum gravity, a number of
different approaches have been proposed, including string theory and
loop quantum gravity as best known examples.
For another prominent approach, see the contributions on causal dynamical
triangulations to this proceedings. The aim of these
models is to provide a microscopic description of quantum spacetime
subsequently to make contact to more macroscopic phenomena, like
e.\,g.~our universe. In doing so it turns out that it is quite hard to
connect the very small length scales on which these models are defined
with the large scales on which observable physics takes
place, e.\,g.~cosmic inflation or particle physics.

A complementary approach towards quantum gravity is to construct
effective theories as an intermediate step between general relativity
and a full theory of quantum gravity and study physical applications
within it. These results can then possibly be used to connect full
quantum gravities to physical phenomena. There have been many
approaches in this direction, from which we choose the approach of
noncommutative (NC) gravity based on a deformation of
symmetries~\cite{Aschieri:2005yw,Aschieri:2005zs} (see also the
review~\cite{Aschieri:2006kc}). For a generalization to NC supergravity, see \cite{vielbein}. 
The main idea behind this formalism is
to replace the classical symmetries of general relativity (i.\,e.~the
diffeomorphisms) by a twist deformed Hopf algebra of diffeomorphisms,
which can be interpreted as quantum symmetries. As a result of this
postulate one obtains that these theories naturally give rise to
noncommutative spacetimes.
Thus we do not deal with a quantized metric field, but 
rather with a metric field on a quantum spacetime.

We will give an introductory overview of the noncommutative gravity theory
developed in  \cite{Aschieri:2005yw,Aschieri:2005zs}. For a deeper discussion we refer
to the original works and the review \cite{Aschieri:2006kc}. After this, we will shortly review 
noncommutative symmetry reduction, discussed previously in \cite{Ohl:2008tw}.
We will keep the discussion very brief and refer to the original paper
for more details. As a next step, we will show under which conditions deformed symmetric models 
solve noncommutative Einstein equations and discuss some phenomenology of deformed cosmological 
and black hole models. More details on exact solutions in NC gravity can be found in \cite{Ohl:2009pv}, 
or in the related works of Schupp and Solodukhin \cite{Schupp:2009pt} and Aschieri 
and Castellani \cite{Aschieri:2009qh}.


\section{\label{sec:ncg}$\star$-products, Hopf algebras, Drinfel'd twists, and all that}
The basic idea of noncommutative geometry is to replace the usual coordinate functions
$x^\mu$ on a manifold by noncommuting operators $\hat x^\mu$, satisfying nontrivial
commutation relations
\begin{flalign}
 \com{\hat x^\mu}{\hat x^\nu} = i\theta^{\mu\nu}(\hat x)~.
\end{flalign}
Analogously to quantum mechanics, the nontrivial commutators imply uncertainty relations
\begin{flalign}
 \Delta x^\mu \Delta x^\nu \neq 0~,
\end{flalign}
leading to an upper bound on the resolution in spacetime.

For our purposes, a more suitable approach to noncommutative geometry is to use $\star$-products instead
of operators. The basic idea is to modify, i.e.~to deform, the point-wise 
product in the algebra of 
functions on the manifold, leading to an associative, but noncommutative product. 
The $\star$-products we will use in this work, the so-called RJS products \cite{Reshetikhin:1990ep,Jambor:2004kc}, are given by
\begin{flalign}
\label{eqn:RJSprod}
 (f \star g)(x) := f(x)~\exp\Bigl(\frac{i\lambda}{2} \overleftarrow{X_\alpha}\Theta^{\alpha\beta}\overrightarrow{X_\beta}\Bigr)~g(x)~,
\end{flalign}
where $X_\alpha\in \Xi$ are mutually commuting vector fields on the manifold satisfying
$\com{X_\alpha}{X_\beta}=0$ and $\Theta^{\alpha\beta}$ is a constant and antisymmetric matrix, which
can be chosen in the canonical (i.e.~Darboux) form. The most widely studied $\star$-product,
the Moyal-Weyl product on $\mathbb{R}^{2n}$, is part of this class. This can be seen by choosing the partial
derivatives as twist generating vector fields, i.e.~$X_\alpha =\partial_\alpha$. Note that
due to the noncommutativity of the $\star$-product, we can have nontrivial commutators as well
\begin{flalign}
 \starcom{f}{g} := f\star g - g\star f\neq 0~.
\end{flalign}

As a next step, we want to discuss the symmetries of theories including $\star$-products.
For this, we have to note that the RJS $\star$-product (\ref{eqn:RJSprod}) can be obtained
by using a so-called (abelian or RJS type) Drinfel'd twist given by
\begin{flalign}
\label{eqn:twist}
 \mathcal{F}=\exp\Bigl(-\frac{i\lambda}{2} \Theta^{\alpha\beta} X_\alpha \otimes X_\beta\Bigr)~.
\end{flalign}
The $\star$-product (\ref{eqn:RJSprod}) is then given by
\begin{flalign}
\label{eqn:RJSprodtw}
 f\star g = \mu\Bigl( \mathcal{F}^{-1} \triangleright f\otimes g \Bigr)~,
\end{flalign}
where $\mu (f\otimes g) = f\cdot g$ is the usual pointwise multiplication map and $\mathcal{F}^{-1}$ is
the inverse twist. Furthermore, $\triangleright$ denotes the action of vector fields on functions
and is defined via the Lie derivative $v\triangleright f:= \mathcal{L}_v(f)=v(f) = v^\mu\partial_\mu f$.
Having identified the twist, we can construct the twisted symmetries of RJS deformed spaces.
In order to do this, we start with the classical infinitesimal diffeomorphisms given by
the Lie algebra of vector fields on the manifold $(\Xi,\com{~}{~})$. From that we
can construct in a canonical way a Hopf algebra $(U\Xi,\cdot,\Delta,S,\epsilon)$,
where $U\Xi$ is the universal envelopping algebra of vector fields, $\cdot$ is the related 
multiplication and $\Delta$ is the coproduct (Leibniz rule). We will not use the antipode 
$S$ and the counit $\epsilon$ in the following and refer to
\cite{Aschieri:2005zs} for a definition. The coproduct is an algebra homomorphism and acts
on the generators as follows
\begin{flalign}
 \Delta(u):= u\otimes 1 + 1\otimes u~,\quad \Delta(1):= 1\otimes 1~.
\end{flalign}

In order to deform the diffeomorphisms, such that they are compatible with the RJS product,
we just have to deform the coproduct as follows
\begin{flalign}
 \Delta_\mathcal{F}(\xi):= \mathcal{F}\Delta(\xi)\mathcal{F}^{-1}~,
\end{flalign}
where $\xi\in U\Xi$. It can be shown that $(U\Xi,\cdot,\Delta_\mathcal{F},S,\epsilon)$
is still a Hopf algebra, but now in general a noncocommutative one. 
We obtain the compatibility condition of the coproduct $\Delta_\mathcal{F}$
\begin{flalign}
 \xi \triangleright (f\star g) = \xi \triangleright \mu_\star(f\otimes g) = \mu_\star\bigl(\Delta_\mathcal{F}(\xi)\triangleright f\otimes g\bigr)~,
\end{flalign}
where $\mu_\star = \mu\circ \mathcal{F}^{-1}$ is the $\star$-multiplication map. 

The $\star$-product (\ref{eqn:RJSprodtw}) can be generalized to $\star$-tensor products, $\star$-wedge products, etc.,~by
 using the corresponding action of vector fields on the involved objects, i.e.~using the
appropriate Lie derivatives \cite{Aschieri:2005zs}.


\section{\label{sec:NCriemann}Basics of noncommutative Riemannian geometry and gravity}
As it was shown in \cite{Aschieri:2005yw,Aschieri:2005zs},
one can construct covariant derivatives and Riemannian curvature on twist deformed spaces. 
These objects are deformed covariant under the Hopf algebra symmetries
 by using suitable combinations of the classical objects and the twist. 

In this article, we will not review the NC Riemannian geometry in a basis free form as presented in \cite{Aschieri:2005zs,Aschieri:2006kc}, but we will give all formulae in a particular simple basis of vector fields and one-forms. This will lead to a better readability of this review article. 

It can be shown that for all RJS twists (\ref{eqn:twist}), which are analytical almost everywhere, we can find a densely defined basis of vector fields $\lbrace e_a\in \Xi:a=1,\dots,n \rbrace$, such that 
\begin{flalign}
\label{eqn:basis}
 \com{e_a}{X_\alpha}=0,\quad\com{e_a}{e_b}=0~,
\end{flalign}
for all $a,b,\alpha$. Here $n$ denotes the dimension of the manifold. This means in particular that the twist acts trivially on the basis vector fields. The associated dual basis $\lbrace \theta^a\rbrace$ defined by the pairing $\pair{e_a}{\theta^b}=\delta^b_a$ has the same property.

Making use of the natural basis (\ref{eqn:basis}), we obtain the following formulae for the geometrical quantities: Given a metric $g=\theta^b\otimes \theta^a g_{ab}$, the inverse metric $g^{-1}=e_a\otimes e_b g^{ba}$ fulfills the condition
\begin{flalign}
 g_{ab}\star g^{ca}= g^{ac}\star g_{ba}= \delta^c_b~,
\end{flalign}
thus it is simply the $\star$-inverse matrix of $g_{ab}$. The associated torsion-free and metric compatible connection, the $\star$-Levi-Civita connection, is given by
\begin{flalign}
\label{eqn:connection}
  \Gamma_{ab}^{~~c}= \frac{1}{2}\bigl(e_a(g_{bd}) + e_b(g_{ad}) -e_d(g_{ab})\bigr)\star g^{cd}~,
\end{flalign}
where $e_a(\cdot)$ denotes the action of the vector field $e_a$ on {\it functions} (i.e.~the Lie derivative on functions).
The $\star$-covariant derivative of a tensor field is given by
\begin{flalign}
(\Nabla^\star_{e_c}\tau)_{b_1\dots b_l}^{a_1\dots a_n}=e_c(\tau^{a_1\dots a_n}_{b_1\dots b_l}) -\Gamma_{cb_1}^{~~\tilde b}\star\tau^{a_1\dots a_n}_{\tilde b\dots b_l} -\dots + \tau^{a_1\dots \tilde a}_{b_1\dots b_l}\star\Gamma_{c \tilde a}^{~~a_n}~.
\end{flalign}
Furthermore, we obtain the $\star$-Riemann tensor
\begin{flalign}
\label{eqn:riemann}
 \mathrm{R}_{abc}^{~~~d}=e_a(\Gamma_{bc}^{~~d}) - e_b(\Gamma_{ac}^{~~d}) +\Gamma_{bc}^{~~e}\star\Gamma_{ae}^{~~d} - \Gamma_{ac}^{~~e}\star\Gamma_{be}^{~~d}~,
\end{flalign}
the $\star$-Ricci tensor $\mathrm{Ric}_{ab}= \mathrm{R}_{cab}^{~~~c}$ and the $\star$-curvature scalar $\mathcal{R}=g^{ab}\star \mathrm{Ric}_{ba}$.
One way of writing the NC dynamics of the metric field are the NC Einstein equations
\begin{flalign}
\label{eqn:einstein}
 G_{ab}:= \mathrm{Ric}_{ab}-\frac{1}{2} g_{ab}\star \mathcal{R} = 8 \pi G_N~T_{ab}~,
\end{flalign}
where we have introduced the NC Einstein tensor $G_{ab}$, a stress-energy tensor $T_{ab}$ and the Newton constant $G_N$. Other possibilities of defining the dynamics of the NC metric field, like e.g.~constructing a deformed version of the Einstein-Hilbert action, will be discussed elsewhere.

In the following, we will construct solutions of the NC Einstein equations (\ref{eqn:einstein}). This is in general a very hard task, since these equations are nonlinear and also nonlocal due to the $\star$-products. As in the classical case, symmetry reduction will turn out to be a helpful tool for constructing symmetric solutions.


\section{\label{sec:symmetryreduction}Noncommutative symmetry reduction}
In classical gravity, it turned out that splitting the problem of describing physical situations into finding background solutions and afterwards perturbing them by field fluctuations on those fixed backgrounds is very fruitful for practical applications. In particular, if the problem obeys some approximate symmetries, like for example isotropy and homogeneity in cosmology, this approach has found many successful applications. Using the example of cosmology, the typical steps are the restriction of the metric and inflaton field to spherical symmetric and translation invariant configurations, leading to the Friedmann equations, which are in general easier to solve than the full Einstein equations. After solving for the background fields, one allows small fluctuations of the metric and scalar field in order to determine, e.g.,~the fluctuations in the cosmic microwave background (CMB).

In this article, we will focus on the first step in the approach described above in a noncommutative setting. The second step, i.e.~the (quantum) field theory of fluctuations, will be discussed elsewhere. To reach our goal, we have to generalize the usual Lie algebra symmetries to quantum Lie algebra symmetries, which will be the basic ingredients of NC symmetry reduction. Furthermore, we will discuss the application of NC symmetry reduction to physical problems, in particular to cosmology and black holes. 

While in classical gravity the symmetries of physical problems can often be described by Lie algebras $(\mathfrak{g},\com{~}{~})$, in noncommutative gravity this mathematical structure has to be replaced by a so-called quantum or $\star$-Lie algebra \cite{Woronowicz:1989rt}. For the twist deformed case, $\star$-Lie algebras can be easily constructed by using the twist \cite{Aschieri:2005zs}. For example, the quantum analogon of the Lie algebra of vector fields $(\Xi,\com{~}{~})$ is given by $(\Xi,\com{~}{~}_\star)$, where the $\star$-Lie bracket $\com{u}{v}_\star = \com{\bar f^\alpha (u)}{\bar f_\alpha(v)}$ is constructed using the commutative Lie bracket and the inverse twist $\mathcal{F}^{-1}=: \bar f^\alpha\otimes \bar f_\alpha$. The same construction can be performed for every classical Lie algebra $(\mathfrak{g},\com{~}{~})$, if the twist is constructed from elements of $\mathfrak{g}$.

In NC gravity, the twist is i.g.~constructed using general vector fields, not necessarily related to the symmetries one wants to implement. This means that we have to look for compatibility conditions among a given classical symmetry Lie algebra $\mathfrak{g}$ and a twist $\mathcal{F}$. For the case of RJS twists (\ref{eqn:twist}), this will reduce to compatibility conditions among the vector fields $X_\alpha$ and the Lie algebra $\mathfrak{g}$. This was done in \cite{Ohl:2008tw} and we found out that the condition
\begin{flalign}
 \com{X_\alpha}{\mathfrak{g}}\subseteq \mathfrak{g},~\forall_\alpha
\end{flalign}
is a necessary and sufficient condition to turn $(\mathfrak{g},\com{~}{~}_\star)$ into a quantum Lie subalgebra of $(\Xi,\com{~}{~}_\star)$. Based on this condition we were able to give a classification of compatible $(\mathcal{F},\mathfrak{g})$-pairs for RJS twists and cosmological and black hole symmetries $\mathfrak{g}$. We refer to \cite{Ohl:2008tw} for the results and only state some models of particular physical interest in section \ref{sec:examples}. This method can also be applied to study deformations of other symmetric spaces, such as for example (anti) de Sitter spaces or black brane scenarios.

Having a quantum Lie algebra symmetry, we have to study the solution of the quantum invariance condition on tensor fields $\tau$ given by
\begin{flalign}
\label{eqn:invariance}
 \mathcal{L}_\mathfrak{g}^\star(\tau) :=\mathcal{L}_{\bar f^\alpha (\mathfrak{g})} \bar f_\alpha(\tau)=\lbrace 0\rbrace~,
\end{flalign}
which is a deformed version of the statement that the symmetries should annihilate symmetric tensors (this is the very definition). In this expression we used the deformed Lie derivative $\mathcal{L}^\star$ constructed by using the twist.
We were able to prove that the quantum invariance condition and the classical one $\mathcal{L}_\mathfrak{g}(\tau)=\lbrace 0\rbrace$ are equivalent, such that the $\star$-invariant tensor fields are just the ones invariant under the classical counterpart of the symmetries. 

In table \ref{tab:comparison} we compare the classical and deformed version of symmetry reduction.

\renewcommand{\arraystretch}{1.5}
\begin{table}
\begin{center}\begin{tabular}{|l|c|c|}
 \hline
~           & cl.~symmetry reduction & NC symmetry reduction   \\ \hline
symmetries  & Lie algebra $(\mathfrak{g},\com{~}{~})$ & $\star$-Lie algebra $(\mathfrak{g},\com{~}{~}_\star)$\\ \hline
restrictions & none      &  $\com{X_\alpha}{\mathfrak{g}}\subseteq \mathfrak{g}~,~\forall_\alpha$\\ \hline
inv.~condition  & $\mathcal{L}_\mathfrak{g}(\tau)=\lbrace 0\rbrace$  & $\mathcal{L}^\star_\mathfrak{g}(\tau)=\lbrace 0\rbrace \Leftrightarrow \mathcal{L}_\mathfrak{g}(\tau)=\lbrace 0\rbrace$ \\ \hline
\end{tabular}\end{center}
\caption{\label{tab:comparison} Comparison between classical (cl.) and NC symmetry reduction.}
\end{table}


\section{\label{sec:dynamics}Dynamics of symmetry reduced sectors: Solutions}
In this section, we will study deformed $\mathfrak{g}$-symmetric solutions of the NC Einstein equations (\ref{eqn:einstein}). 
For this we use the framework explained in section \ref{sec:symmetryreduction}, see in particular table \ref{tab:comparison}, 
and use as ansatz $\star$-symmetric tensor fields satisfying $\mathcal{L}_\mathfrak{g}^\star(\tau)=\lbrace0\rbrace$. This results in consistently 
deformed equations for the symmetry reduced tensor fields, like e.g.~deformed Friedmann equations in cosmology.

As it was recognized and explained in \cite{Ohl:2009pv}, as well as in the related works \cite{Schupp:2009pt,Aschieri:2009qh},
the deformed symmetric equations of motion reduce to the undeformed ones, if the twist (\ref{eqn:twist}) satisfies
\begin{flalign}
\label{eqn:semikilling}
 \Theta^{\alpha\beta} X_\alpha\otimes X_\beta ~\subseteq ~\Xi\otimes\mathfrak{g} + \mathfrak{g} \otimes \Xi~.
\end{flalign}
Twists of this form are called semi-Killing, since using the canonical form of $\Theta^{\alpha\beta}$,
half of the vector fields $X_\alpha$ have to be Killing vector fields, i.e.~$X_\alpha\in\mathfrak{g}$.
The fact that all $\star$-products between $\star$-symmetric tensor fields drop out can be easily understood, 
since acting with (\ref{eqn:semikilling}) on $\star$-symmetric tensor fields is trivial due to their remaining classical symmetries $\mathcal{L}_\mathfrak{g}(\tau)=\lbrace0\rbrace$, see table \ref{tab:comparison}.

Since, at least in principle, we know how to solve the classical equations of Einstein gravity coupled to matter, this leads to solutions of the NC
Einstein equations (\ref{eqn:einstein}). Exact solutions beyond the semi-Killing case are more complicated, but we have found one particular example in \cite{Ohl:2009pv}, see also section \ref{sec:examples}. Furthermore, in \cite{Aschieri:2009qh} affine Killing twists have been considered, leading to another class of exactly solvable models.

As a last note, we want to add that the reduction of the deformed symmetry reduced dynamics to the undeformed one in the semi-Killing case does not mean that these models are trivial. In contrast, these models will indeed receive distinct NC effects from field fluctuations on the background solutions, as well as from uncertainties of the coordinate operators.


\section{\label{sec:examples}Examples of deformed cosmological and black hole solutions}
In this section we will give some examples of deformed cosmological (Friedmann-Robertson-Walker) models and deformed Schwarzschild black holes. 
For a complete classification of these models see \cite{Ohl:2008tw}. Furthermore, see \cite{Ohl:2009pv} for a deeper discussion of what follows.

Consider the following families of deformed FRW models $\mathfrak{C}_{AB}$ and black hole models $\mathfrak{B}_{AB}$:
\begin{subequations}
\label{eqn:models}
\begin{flalign}
\label{eqn:c22}
 &\mathfrak{C}_{22}:\qquad X_1=X_1^0(t)\partial_t~,\quad X_2=d\partial_\phi + f_2 r\partial_r~,\\
\label{eqn:c32}
 &\mathfrak{C}_{32}:\qquad X_1=X_1^0(t)\partial_t+d \partial_\phi~,\quad X_2=X_2^0(t)\partial_t + f_2 r\partial_r~,\\
\label{eqn:b12}
&\mathfrak{B}_{12}:\qquad X_1 = c_1^0 \partial_t +\kappa_1\partial_\phi~,\quad X_2 =c_2^0(r) \partial_t + \kappa_2 \partial_\phi +f(r) \partial_r~.
\end{flalign}
\end{subequations}
Here we used (comoving) spherical coordinates $(r,\zeta,\phi)$, as well as some freely chooseable parameter(function)s.
These models are examples within the classification of admissible $(\mathcal{F},\mathfrak{g})$-pairs given in \cite{Ohl:2008tw}.
The associated $\star$-commutation relations among appropriate coordinate functions are given by
\begin{subequations}
\label{eqn:commutators}
\begin{align}
\label{eqn:a22}
  \mathfrak{C}_{22}:\; &
    \left\{ \begin{aligned}
               \starcom{t}{\exp i\phi}&=-2\exp i\phi~\sinh\Bigl(\frac{\lambda d}{2} X_1^0(t)\partial_t\Bigr) t\\
               \starcom{t}{r}         &=2 i r ~\sin\Bigl(\frac{\lambda f_2}{2} X_1^0(t)\partial_t\Bigr)t
            \end{aligned} \right. \\
\label{eqn:a32}
  \mathfrak{C}_{32}:\; &
    \left\{ \begin{aligned}
               \starcom{t}{\exp i\phi}&= 2 \exp i\phi~\sinh\Bigl(\frac{\lambda d}{2} X_2^0(t)\partial_t\Bigr) t\\
               \starcom{t}{r}         &=2 i r ~\sin\Bigl(\frac{\lambda f_2}{2} X_1^0(t)\partial_t\Bigr)t\\
               \exp i\phi\star r      &= e^{-\lambda d f_2}~r\star \exp i\phi
    \end{aligned} \right. \\
\label{eqn:a12}
  \mathfrak{B}_{12}:\; &
     \left\{ \begin{aligned}
               \starcom{t}{\exp i\phi} &=\exp i\phi~ \Bigl(2 \sinh\Bigl(\frac{\lambda\kappa_1}{2}\bigl(c_2^0(r)\partial_t
                                          + f(r)\partial_r\bigr)\Bigr) t -\lambda \kappa_2 c_1^0\Bigr)\\
               \starcom{t}{r}          &=i\lambda c_1^0 f(r)~,\\
               \starcom{\exp i\phi}{r} &=-2 \exp i\phi ~\sinh\Bigl(\frac{\lambda\kappa_1}{2}f(r)\partial_r\Bigr)r
     \end{aligned} \right..
\end{align}
\end{subequations}

It turns out that these models are semi-Killing for the choice $f_2=0$ in model $\mathfrak{C}_{22}$, $X_1^0(t)\equiv 0$ in model $\mathfrak{C}_{32}$, as well as for all choices of parameters in model $\mathfrak{B}_{12}$. This leads to exact solutions describing NC cosmologies and black holes.

Next, we will discuss physical implications of the nontrivial coordinate algebras for some of our models. Consider for example the model $\mathfrak{C}_{22}$
(\ref{eqn:c22}) with $f_2=0$ and for simplicity $X_1^0(t)\equiv1$. Then the coordinate algebra (\ref{eqn:a22}) 
reduces to the algebra of a quantum mechanical particle on the circle, i.\,e.
\begin{flalign}
 \com{\hat E}{\hat t}=\lambda \hat E~,
\end{flalign}
where we introduced the abstract operators $\hat t$ and $\hat E:=\widehat{\exp i\phi}$ and set $d=1$. This algebra previously appeared e.\,g.~in
 the context of noncommutative field theory~\cite{ncfield} and the noncommutative BTZ black hole~\cite{Dolan:2006hv}.
 It is well known that the operator $\hat t$ can be represented as a differential operator acting on the Hilbert
space $L^2(S_1)$ of square integrable functions on the circle and the spectrum can be shown to be given by $\sigma(\hat t)=\lambda (\mathbb{Z}+\delta)$, where $\delta\in[0,1)$ labels unitary inequivalent representations. 
The spectrum should be interpreted as possible time eigenvalues.  
Thus our model with discrete time can be used to realize singularity avoidance
in cosmology.
Consider for example an
inflationary background with scale factor $A(t)=t^p$, where $p>1$ is a parameter. This so-called power-law inflation can be 
realized by coupling a scalar field with exponential potential to the geometry even in our NC model, since the symmetry reduced Riemannian geometry reduces to the undeformed one as explained above. Note that the scale factor goes to zero at the time $t=0$ and leads
to a singularity in the curvature scalar. But as we discussed above, the possible time eigenvalues are $\lambda (\mathbb{Z}+\delta)$,
 which for $\delta\neq0$ does not include the time $t=0$.

Taking a look at the coordinate algebra of the black hole (\ref{eqn:a12}), we observe that it includes in particular the
algebra of a quantum mechanical particle on the circle for time and angle variable, if we choose $c_2^0(r)\equiv0$ and 
$f(r)\equiv 0$. This leads to discrete times.
Another simple choice is $c_1^0=\kappa_2=0$, $c_2^0(r)\equiv0$, $\kappa_1=1$ and $f(r)=r$. The radius spectrum in this case is 
$\sigma(\hat r)=\Lambda\exp\bigl(\lambda(\mathbb{Z}+\delta) \bigr)$, describing a fine grained geometry around the black hole.
The phenomenological problem with this model is that the spacings between the radius eigenvalues grow exponentially in $r$.
This can be fixed by considering a modified twist like e.\,g.~$c_1^0=\kappa_2=0$, $c_2^0(r)\equiv0$, $\kappa_1=1$ and $f(r)=\tanh \frac{r}{\Lambda}$, where $\Lambda$ is some length scale.
The essential modification is to choose a bounded $f(r)$. 
Consider the coordinate change $r\to \eta=\log \sinh (\frac{r}{\Lambda})$, then the
algebra (\ref{eqn:a12}) in terms of $\eta$ becomes
\begin{flalign}
 \com{\hat E}{\hat \eta}=-\frac{\lambda}{\Lambda} \hat E~,
\end{flalign}
leading to the spectrum $\sigma(\hat \eta)=\frac{\lambda}{\Lambda} \bigl(\mathbb{Z}+\delta\bigr)$.
The spectrum of $\hat r$ is then given by $\sigma(\hat r) = \Lambda~ \mathrm{arcsinh}\exp\bigl(\frac{\lambda}{\Lambda}(\mathbb{Z}+\delta)\bigr) $. This spectrum approaches constant spacings
 between the eigenvalues for large $r$.

We will close this section by giving an example of a non-semi-Killing solution of the NC Einstein equations (\ref{eqn:einstein}).
Consider the model $\mathfrak{C}_{22}$ with $d=0$ (\ref{eqn:c22}).
It turns out that both $X_\alpha\not\in\mathfrak{g}$, therefore the Riemannian geometry does not to reduce to the undeformed one. 
Thus we expect corrections in $\lambda$ to the NC Einstein equations (\ref{eqn:einstein}) and its solutions.

Consider now the (undeformed) de Sitter space given by the scale factor $A(t)=\exp H t$, where $H$ is the Hubble parameter.
It turns out that all $\star$-products entering the deformed geometrical quantities (see section~\ref{sec:NCriemann})
reduce to the undeformed ones, if $X_1^0(t)\equiv1$. Thus the undeformed de Sitter space solves NC Einstein equations (\ref{eqn:einstein}) 
for this particular choice of twist and an undeformed cosmological constant. Note that in contrast to the solutions above, 
we required the explicit form of the scale factor $A(t)$.


\section{Conclusions and outlook}
In this paper we gave an introductory overview of the noncommutative gravity theory defined in \cite{Aschieri:2005yw,Aschieri:2005zs}.
This theory is constructed in such a way that it is covariant under a deformed diffeomorphism symmetry, mathematically described by 
a twist deformed Hopf algebra of diffeomorphisms.
In section \ref{sec:symmetryreduction} we reviewed our approach to noncommutative symmetry reduction \cite{Ohl:2008tw}. We emphasized
the role of quantum Lie algebras and constructed compatibility conditions among the twist $\mathcal{F}$ 
and the desired symmetry $\mathfrak{g}$. Based on this approach, we discussed the issue of solvability of 
the noncommutative Einstein equations in section \ref{sec:dynamics}. We found out under which conditions the NC Einstein 
equations reduce to the undeformed ones, thus simplifying the search for exact solutions.
We gave explicit examples of quantum Lie algebra symmetric
cosmological and black hole models solving the noncommutative Einstein equations in section \ref{sec:examples}.

In a future work \cite{OhlSchenkel:fluct}, we will investigate classical and quantum field fluctuations on background
solutions. This is a step required in order to make contact to physics, such as for example the cosmic microwave background (CMB).
In order to formulate and quantize field theories in a deformed covariant way, we will use deformed Poisson algebras
 as defined in \cite{poisson}.

\section*{Acknowledgments}
I want to thank the organizers and participants of the 49.~Cracow School of Theoretical Physics 2009 in Zakopane 
for this very interesting conference.
Furthermore, I want to thank the faculty of physics and astronomy of the university of W\"urzburg for the financial 
support through the ``Wilhelm-Conrad-R\"ontgen-Studienpreis''.
This research is supported by Deutsche
Forschungsgemeinschaft through the Research Training Group 1147
\textit{Theoretical Astrophysics and Particle Physics}.


\end{document}